\documentclass[journal=jctcce,manuscript=article]{achemso}

\setkeys{acs}{doi=true}

\usepackage[T1]{fontenc}
\usepackage{xcolor}
\usepackage{soul}
\usepackage{amsfonts}
\usepackage{amssymb}
\usepackage{amsmath}
\usepackage{bm}
\usepackage{lmodern}
\usepackage{mathtools}
\usepackage[colorlinks=true,linkcolor=blue,citecolor=blue,urlcolor=blue]{hyperref}
\usepackage{tikz}
\usetikzlibrary{arrows.meta}

\DeclarePairedDelimiter\ppar{(}{)}              % ( )
\DeclarePairedDelimiter\pnrm{\lVert}{\rVert}    % || ||
\DeclarePairedDelimiter\pset{\{}{\}}            % { }

\newcommand{\rtab}[1]{Tab.~\ref{#1}}
\newcommand{\rfig}[1]{Fig.~\ref{#1}}
\newcommand{\rsfig}[2]{Fig.~\ref{#1}\,#2}

\newcommand{\rmfig}[1]{Fig.~{\color{blue}{S{#1}}}}

\newcommand{\rsct}[1]{Sec.~\ref{#1}}

\newcommand{\req}[1]{Eq.~\ref{#1}}

\newcommand{\dd}[1]{\operatorname{d#1}}

\newcommand{\bz}{\mathbf{z}}

\newcommand{\bx}{\mathbf{x}}

\newcommand{\dz}{\dd{\mathbf{z}}}

\newcommand{\e}{\operatorname{e}}
\newcommand{\kT}{k_{\mathrm{B}}T}
\newcommand{\kb}{k_{\mathrm{B}}}

\author{Mariia Ivonina}
\email{m.ivonina@kyudai.jp}
\affiliation{%
    Platform of Inter/Transdisciplinary Energy Research, Kyushu University, 744, Motooka, Nishi-ku, Fukuoka 819-0395, Japan
}
\altaffiliation{%
    Current address: RIKEN Center for Interdisciplinary Theoretical and Mathematical Sciences (iTHEMS), RIKEN, Wako 351-0198, Japan
}

\author{Jakub Rydzewski}
\affiliation{%
  Institute of Physics,
  Faculty of Physics, Astronomy and Informatics,
  Nicolaus Copernicus University,
  Grudzi\k{a}dzka 5, 87-100 Toru\'n, Poland
}
\email{jr@fizyka.umk.pl}

\title{\large Unraveling the Mechanism of Drug Binding to SARS-CoV-2 RNA
Pseudoknot with Thermodynamics-Driven Machine Learning}

\begin{document}

\clearpage

\begin{abstract}
The pseudoknot secondary structure in SARS-CoV-2 RNA is essential for regulating protein synthesis through $-$1 programmed ribosomal frameshifting ($-1$ PRF), a mechanism that allows the virus to generate both structural and non-structural proteins from overlapping reading frames. This pseudoknot exhibits both threaded and unthreaded long-lived topologies. The influence of ligand binding on its folding is a process critical for the development of $-$1 PRF small-molecule inhibitors. Understanding this process through unbiased molecular dynamics (MD) simulations can be facilitated by introducing collective variables (CVs) that capture the corresponding slowest dynamical modes. Here, we use spectral map (SM), a thermodynamics-driven machine learning technique, to learn such CVs directly from all-atom MD trajectories of the SARS-CoV-2 RNA pseudoknot in complex with the $-$1 PRF inhibitor merafloxacin and its two structural analogs in neutral and ionized forms. Free-energy landscapes (FELs) derived from the learned CVs indicate that ligand-induced destabilization is topology-selective. In the threaded pseudoknot, the inhibitors destabilize the S2 stem, while in the unthreaded pseudoknot, destabilization occurs in the S1 and S3 stems. Furthermore, the extent to which each ligand reshapes the FEL matches experimentally reported antiviral potency, whereas the protonation state qualitatively alters dynamics within the same RNA topology. Overall, our results show how pseudoknot topology, ligand type, and protonation state collectively influence the slow conformational dynamics of viral RNA and establish physiological protonation as a critical factor for modeling RNA-targeted drug action.
\end{abstract}

\maketitle

\clearpage

\section{Introduction}
\label{sec:into}
Programmed $-1$ ribosomal frameshifting ($-1$ PRF) is an essential regulatory mechanism used by many RNA viruses, including coronaviruses and HIV, to control the synthesis of proteins required for viral replication~\cite{dinman2012mechanisms}. In this process, the translating ribosome slips backward by one nucleotide on the viral RNA and enters an alternative reading frame, altering the encoded codons. An efficient $-1$ PRF is required to maintain the correct balance of viral proteins produced, making the RNA elements responsible for frameshifting particularly attractive antiviral targets. The frameshifting event is typically stimulated by two cis-acting RNA elements: a heptanucleotide slippery sequence of the form X XXY YYZ and a downstream RNA pseudoknot. The pseudoknot acts as a mechanical roadblock that pauses ribosomal elongation and increases the probability of ribosomal shift into the $-1$ frame at the slippery sequence. Because the efficiency of this process depends on specific RNA conformations, the pseudoknot emerges as an appealing target for antiviral drug development. Unlike protein-targeting strategies that intervene after viral gene expression has begun, targeting viral RNA can suppress propagation at an earlier stage, before viral proteins are fully produced and assembled~\cite{falese2021targeting,childs2022targeting,kovachka2024small,rodriguez2025electrostatic}.

Following the outbreak of severe acute respiratory syndrome coronavirus 2 (SARS-CoV-2) in late 2019, substantial effort has been devoted to developing vaccines and antiviral therapies to reduce viral transmission and mortality. In SARS-CoV-2, $-1$ PRF is estimated to occur in more than 50\% of all translation events~\cite{finkel2021coding}, making the frameshift-stimulatory pseudoknot a particularly promising therapeutic target and highlighting the need to understand the structural basis of its function. Experimental techniques such as X-ray crystallography~\cite{jones2022crystal,roman2021sars}, cryo-EM~\cite{zhang2021cryo,bhatt2021structural,peterson2024structure} and NMR~\cite{wacker2020secondary,neupane2021structural,pekarek2023cis}, as well as molecular dynamics (MD) simulations~\cite{omar2021modeling,schlick2021structure,yan2022length,he2023atomistic}, have captured the conformational heterogeneity of the pseudoknot in solution, showing that the SARS-CoV-2 frameshift signal adopts a three-stem H-type pseudoknot that can populate multiple conformers, including 5$'$-end threaded and unthreaded topologies. Overall, this indicates that, while the pseudoknot retains a relatively stable secondary structure, it can undergo substantial tertiary rearrangements.

Several inhibitors that affect $-1$ PRF in coronaviruses via interaction with the pseudoknot have been identified, demonstrating the feasibility of RNA-targeted antiviral modulation~\cite{park2011identification}. MTDB, a small-molecule inhibitor, retains activity against virus replication for both SARS-CoV-1 and SARS-CoV-2~\cite {kelly2020structural,neupane2020anti,munshi2022identifying}. Merafloxacin, a fluoroquinolone antibacterial, remains an effective $-1$ PRF inhibitor against naturally occurring SARS-CoV-2 mutants~\cite{sun2021restriction}. Nafamostat, a serine protease inhibitor, demonstrates broad-spectrum activity against multiple coronavirus-derived frameshift signals, with effects comparable to those of merafloxacin~\cite{zafferani2022rt,munshi2022identifying}. Geneticin, an aminoglycoside antibiotic, exhibits antiviral activity against multiple SARS-CoV-2 variants~\cite{Varricchio2022GeneticinSS}. Similarly, aminoquinazoline derivatives act as inhibitors of $-1$ PRF by interacting with the SARS-CoV-2 pseudoknot structure. Additional compounds with similar pseudoknot-targeting mechanisms have also been identified~\cite{chen2020drug,ahn2021novel,mathez2024novel}.

Despite this progress, the dynamical basis of ligand recognition by the SARS-CoV-2 pseudoknot remains incompletely understood. Prior computational results have shown that ligand binding can strongly depend on RNA topology. For example, $5'$-end threading can create clefts that are absent in unthreaded conformation~\cite{kelly2020structural}, and some ligands remain bound to the pseudoknot on the nanosecond timescale~\cite{mathez2024novel}. Nevertheless, the mechanism remains unresolved: it is still unclear whether a ligand stabilizes a unique binding mode, migrates among multiple transient interaction sites, or actively reshapes the RNA conformation. 

Unlike protein binding pockets~\cite{rydzewski2017ligand,bernetti2019kinetics}, RNA binding sites are unstructured, shallow, transient, and topology-dependent~\cite{warner2018principles,veenbaas2025ligand}. Therefore, detailed descriptions of ligand recognition require quantifying equilibrium thermodynamic properties, such as free-energy landscapes (FELs), and assessing how these landscapes are altered by inhibition~\cite{valsson2016enhancing,bussi2020using,sega2026molecular}.

A further consideration specific to RNA-targeted ligands is the charge state of the bound compound. Because the RNA backbone is polyanionic, electrostatic contacts with positively charged or zwitterionic moieties often dominate recognition and can alter affinity, selectivity, and association kinetics by orders of magnitude~\cite{rodriguez2025electrostatic}. For example, this is possible for merafloxacin: like other fluoroquinolones, it carries both an acidic carboxylate and a basic amine and is predominantly zwitterionic at physiological pH~\cite{van2014fluoroquinolone}, suggesting that its inhibitory functions might be associated with strong electrostatic effects.

Conformational changes relevant to RNA function span a broad range of timescales, from picoseconds for puckering and bond rotations, to nanoseconds for base-pair opening, and to much longer timescales for large-scale conformational switching~\cite{sponer2018rna,mlynsky2018exploring,bernetti2023integrating,languin2026rna}. Experimental methods provide only limited access to this microscopic dynamics, yet such information is essential for a mechanistic understanding of RNA-mediated processes. MD simulations offer an atomistic view of RNA motions. However, quantitatively characterizing the thermodynamics and kinetics of RNA conformational transitions requires collective variables (CVs) that can identify and disentangle slow dynamical modes~\cite{peters2016reaction,rogal2021reaction,rydzewski2023manifold,gokdemir2025machine,zhu2025enhanced}. For RNA folding, CVs must distinguish between long-lived stem-connected and stem-disrupted states, locate transition regions, and capture motions associated with barriers of several $k_{\mathrm{B}}T$.

To address this challenge, we employ a thermodynamics-driven machine learning (ML) framework to construct CVs for SARS-CoV-2 pseudoknot folding. Specifically, we use spectral map (SM)~\cite{rydzewski2023spectral,rydzewski2024learning,rydzewski2024spectral}, a recently developed method for performing dimensionality reduction using neural networks (NNs) and constructing CVs without prior knowledge of the underlying process. SM constructs CVs from thermodynamic information, yielding reaction coordinates that correspond to the slow modes of the system. For a detailed discussion regarding physics-based ML methods for identifying CVs, we refer to recent reviews by G\"okdemir and Rydzewski~\cite{gokdemir2025machine}, and Zhu et al~\cite{zhu2025enhanced}. This characteristic distinguishes SM from other ML techniques for learning CVs, making SM especially useful for characterizing ligand-induced conformational transitions in RNA pseudoknots, where the coupling between ligand association and large-scale RNA rearrangement can benefit from a physics-based approach.

In this work, we investigate how the SARS-CoV-2 frameshift-stimulatory pseudoknot interacts with the $-1$ PRF inhibitor merafloxacin and two structural analogs reported by Sun et al~\cite{sun2021restriction}. We analyze how ligand structure, protonation state, and pseudoknot topology jointly determine RNA structural stability and drive conformational transitions. To achieve this, we perform extensive all-atom MD simulations of both threaded and unthreaded pseudoknot models in ligand-free and ligand-bound states, and analyze the resulting trajectories with SM to identify the slowest modes and trace the minimum free-energy pathways (MFEPs) in the FELs of each RNA--ligand system.

Our results show that ligand-induced destabilization is topology-selective. The threaded topology predominantly loses native contacts at the S2 stem, whereas the unthreaded topology disrupts native contacts in the S1 and S3 stems. The SM-learned slow modes reveal a clear dynamical distinction between the two topologies: merafloxacin zwitterion binding imposes structure on the otherwise featureless unthreaded pseudoknot, indicating that the inhibitor-induced dynamics resemble those of a threaded pseudoknot. Next, we find that how each ligand reshapes the learned FEL agrees with the experimentally reported antiviral ordering~\cite{sun2021restriction}. Finally, our results demonstrate that the neutral and zwitterionic forms of merafloxacin yield qualitatively different FELs on the same RNA topology, confirming that physiological protonation is crucial to the mechanistic picture of RNA--ligand recognition.

\section{Thermodynamics-Driven Machine Learning}

\subsection{Dimensionality Reduction with an Encoder}
\begin{figure}[ht]
  \input{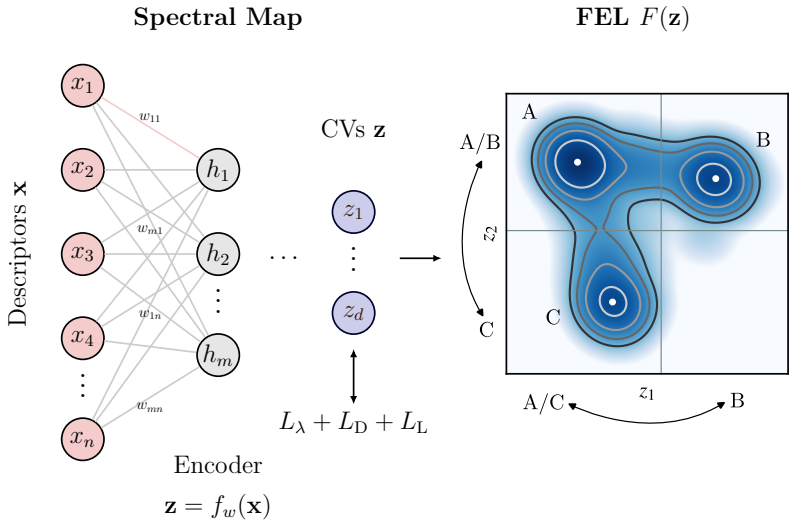}
  \caption{%
    Spectral map (SM) outline. The encoder $f_\theta$ maps high-dimensional descriptors $\bx$ to a low-dimensional latent space of CVs $\bz$. The total loss combines three terms: $L_\lambda$ (spectral loss to capture slow dynamics), $L_{\mathrm{D}}$ (decorrelation loss to ensure CVs represent distinct modes), and $L_{\mathrm{L}}$ (graph Laplacian regularizer for smoothness). The learned CVs $\bz$ are used to estimate the free-energy landscape (FEL) $F(\bz)$. Then, the CVs can be used to describe transitions between metastable states (A, B, and C): $z_1$ between A/C and B, and $z_2$ between A/B and C. These transitions are tied to the slowest dynamic modes in the system.
  }
  \label{fig:diagram}
\end{figure}

SM employs an NN encoder $f_\theta(\bx)$ that learns slow CVs directly from MD data~\cite{rydzewski2023spectral,rydzewski2024learning,rydzewski2024spectral}, capturing nonlinear relationships in the data by optimizing for dynamical slowness of the learned CVs. The encoder $f_\theta: \mathbb{R}^n \to \mathbb{R}^d$ (see \rfig{fig:diagram}) maps high-dimensional descriptors $\bx = (x_1, \dots, x_n)$ to a reduced space of $d$ CVs, $\bz = f_\theta(\bx) = (z_1, \dots, z_d)$, where $d \ll n$ and $\theta$ denotes the NN weights.

\subsection{Markov Transition Matrix}
To learn slow CVs, we train the encoder by maximizing the separation between the slowest (dominant in magnitude) and fastest eigenvalues of a Markov transition matrix $M(\bz_k,\bz_l)$ constructed in $\bz$ space. The algorithm to construct $M$ proceeds as  follows~\cite{rydzewski2023spectral,rydzewski2024learning,rydzewski2024spectral}:
\begin{enumerate}
  \item We compute a Gaussian kernel $G(\bz_k,\bz_l)=\exp\ppar*{-\frac{1}{\varepsilon}{\pnrm{\bz_k-\bz_l}^2}}$, where $\pnrm{\bz_k-\bz_l}^2$ denotes the squared Euclidean distance between CV samples and $\varepsilon$ is the bandwidth parameter.

  \item To correct for non-uniform sampling density, we employ an anisotropic diffusion kernel~\cite{nadler2006diffusion}:
    \begin{equation}
      \label{eq:kernel}
      K(\bz_k,\bz_l) = \frac{G(\bz_k,\bz_l)}{\sqrt{\varrho(\bz_k)\varrho(\bz_l)}},
    \end{equation}
    where $\varrho(\bz_k)=\sum_l G(\bz_k,\bz_l)$ is a kernel density estimate.

  \item We compute the Markov transition matrix $M$ by row-normalizing $K$:
    \begin{equation}
      \label{eq:markov}
      M(\bz_k,\bz_l) = D(\bz_k)^{-1} K(\bz_k,\bz_l),
    \end{equation}
    where $D(\bz_k) = \sum_n K(\bz_k,\bz_n)$. The matrix $M$ models a discrete Markov chain in the CV space, which for a large sample size and small bandwidth, converges to the Fokker--Planck operator associated with the equilibrium ensemble~\cite{coifman2006diffusion,nadler2006diffusion}, $p(\bx) \propto \exp(-\beta U(\bx))$ with the underlying potential $U(\bx)$.
    
  \item We compute the symmetric conjugate $M_{\mathrm{s}} = D^{1/2} M D^{-1/2}$. This similarity transformation preserves the eigenvalues of $M$ while yielding a symmetric matrix amenable to symmetric eigendecomposition.

  \item We perform an eigendecomposition $M_{\mathrm{s}}\Phi=\lambda\Phi$, obtaining eigenvalues $\lambda_0 = 1 > \lambda_1 \ge \dots \ge \lambda_{N-1}$, which approximate the slowest timescales of the system. We use these eigenvalues are used to define the spectral loss.
\end{enumerate}

\subsection{Spectral Loss}
\label{ssec:loss}
The transition matrix is then used to define a loss function that SM optimizes during training to obtain slow CVs. In this work, we use the negative of cumulative spectral weight of the $k$ leading eigenvalues as the main loss:
\begin{equation}
  \label{eq:loss-eig}
  L_\lambda = %
    -\frac{\sum_{i=0}^{k-1} \lambda_{i}}{\sum_{i=0}^{N-1} \lambda_i},
\end{equation}
where $N$ is the batch size. As the dominant eigenvalues of $M$ correspond to the longest timescales in the system dynamics, this loss term ensures that the slowest degrees of freedom are captured by the encoder. Unlike the spectral gap $\Delta\lambda = \lambda_{k-1} - \lambda_k$ used in our previous work~\cite{rydzewski2023spectral,rydzewski2024learning,rydzewski2024spectral}, the cumulative spectral weight is less sensitive to the choice of $k$. An incorrect guess for the number of metastable states can significantly affect CVs learned via the spectral gap, whereas the cumulative weight provides a smoother loss. For a detailed discussion, we refer to Sec.~S1 in SI.

In addition to $L_\lambda$, we employ two auxiliary losses. First, we use a decorrelation loss:
\begin{equation}
  \label{eq:loss-decorr}
  L_{\mathrm{D}} = \frac{1}{d(d-1)} \sum_{k \neq l} R^2_{kl},
\end{equation}
where $R_{kl}$ is the correlation coefficient between CVs $z_k$ and $z_l$, and the factor $d(d-1)$ counts the off-diagonal pairs. This loss encourages decorrelation among the learned CVs, ensuring that each CV represents a distinct slow mode. Second, a graph Laplacian regularizer is added:
\begin{equation}
  \label{eq:loss-lapl}
  L_{\mathrm{L}} = %
    \frac{1}{2} \sum_{kl} M_{\mathrm{s}}(\bz_k, \bz_l) \pnrm*{\bz_k - \bz_l}^2.
\end{equation}
This term penalizes large differences in CV values between states with high transition probability, enforcing smoothness of the learned representation with respect to the underlying dynamics. The total loss combines these three components: $L = L_\lambda + L_{\mathrm{D}} + L_{\mathrm{L}}$.

\subsection{Orthogonality}
To stabilize training, we enforce orthogonality on the weight matrices in the hidden layers. This constraint prevents gradient explosion during backpropagation and ensures that the NN does not collapse to trivial solutions. With the orthogonality constraints, the encoder is forced to preserve the scale of its representation, which directly affects the spectral properties of the learned CVs. The training is focused on arranging CV space so that a few leading eigenvalues are genuinely large (reflecting slow dynamics), rather than trivially collapsing the scale to reach the minimum of the spectral loss. Specifically, for a weight matrix $W \in \mathbb{R}^{m \times n}$ connecting neighboring linear layers with $n$ inputs and $m$ outputs, we impose $W^\top W = \mathrm{Id}_n$ if $m \geq n$ (tall matrix) and $W W^\top = \mathrm{Id}_m$ if $m < n$ (wide matrix), where $\mathrm{Id}$ denotes an identity matrix. The orthogonality is enforced using the Cayley parametrization via Householder reflectors~\cite{lezcano_casado2019trivializations}.

\section{Computational Details}
\subsection{RNA Models and Ligands}
\begin{figure}
  \includegraphics[width=\textwidth]{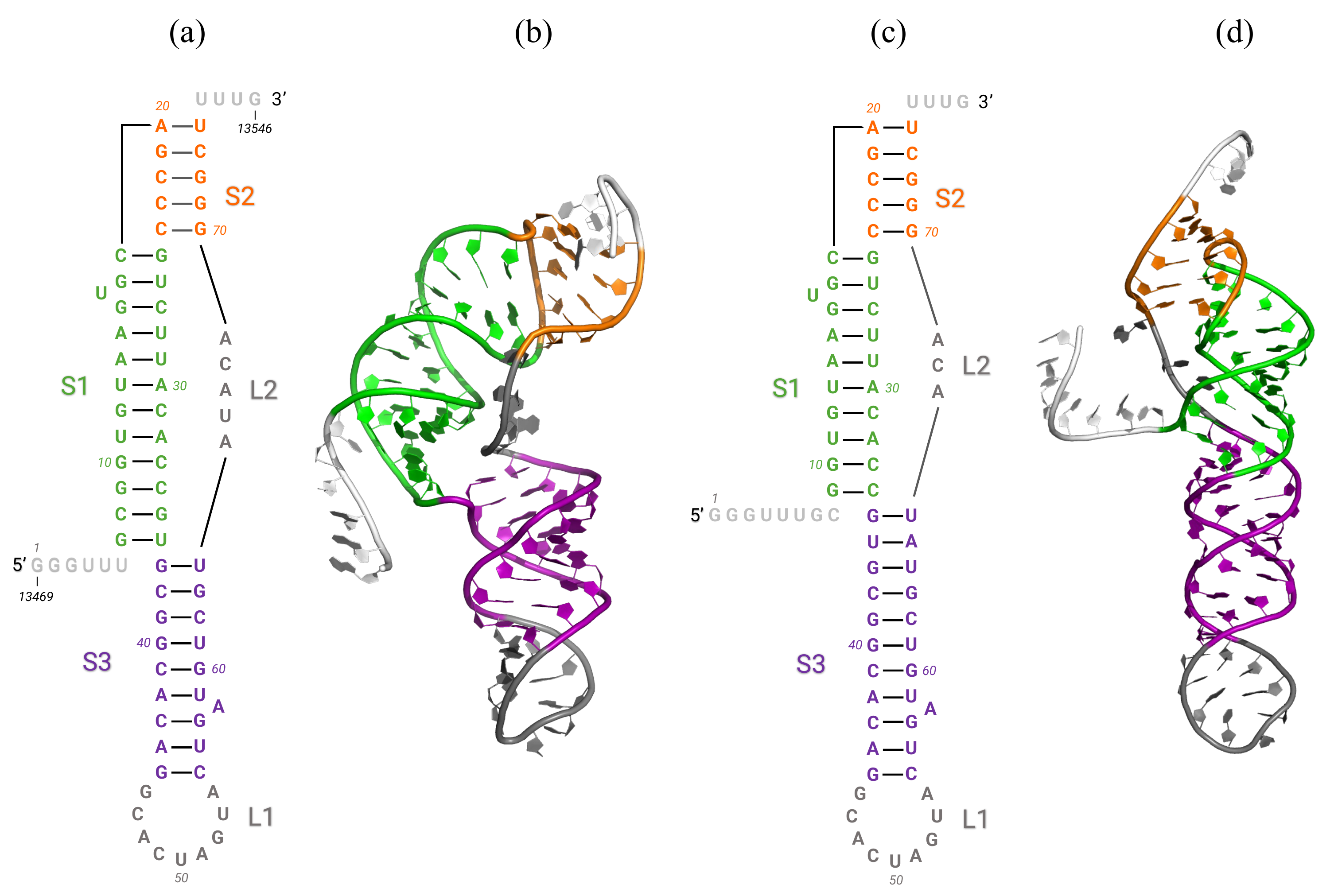}
  \caption{%
    Secondary structures of the SARS-CoV-2 $-1$ ribosomal frameshift-stimulatory pseudoknot derived from crystallographic data. (a) Threaded (RT) conformation~\cite{zhang2021cryo}, in which the 5$'$-end passes through the junction formed by stems 1 (S1), 2 (S2), and 3 (S3), producing a topological tertiary fold. (b) 3D representation of the RT model. (c) Unthreaded (RU) conformation~\cite{peterson2024structure}, in which this topological fold is absent. (d) 3D representation of the RU model. Nucleotides are colored according to their secondary-structure elements (stems: S1 in green; S2 in orange; S3 in purple; and loops L1, L2 in dark gray). The 5$'$ and 3$'$ termini are depicted in light gray.
  }
  \label{fig:rt-ru-2d-3d}
\end{figure}

The initial structures of the SARS-CoV-2 pseudoknot used for MD simulations are retrieved from the Nucleic Acid Knowledgebase (NAKB) (\url{https://nakb.org/}). We use two cryo-EM pseudoknot structures, 6XRZ~\cite{zhang2021cryo} and 8VCI~\cite{peterson2024structure}. These two models differ in their fold topology: in 6XRZ, the 5$'$-end RNA strand is threaded through the junction between stem 1 (S1) and stem 3 (S3), whereas in 8VCI the 5$'$-end remains outside this junction (\rfig{fig:rt-ru-2d-3d}). Hereafter, we refer to these RNA conformers as the threaded (RT) and unthreaded (RU) forms.

To make both conformers comparable in this study, we extract 67 nucleotides from 6XRZ and 8VCI, corresponding to the main elements of the pseudoknot: stems 1, 2, and 3 (S1, S2, S3) and loops 1 and 2 (L1, L2). To preserve the native folding of the pseudoknot at its boundaries, we include seven nucleotides upstream at the 5$'$-end (GGGUUU) and four nucleotides downstream at the 3$'$-end (UUUG). In total, 78 nucleotides that constitute the SARS-CoV-2 pseudoknot body in both threaded and unthreaded conformations, as shown in \rfig{fig:rt-ru-2d-3d}, are simulated in this study.

To investigate ligand recognition by the SARS-CoV-2 pseudoknot, we focus on antibiotics from the fluoroquinolone family. These compounds share a characteristic core structure comprising a bicyclic ring system bearing a carboxylic acid group at the 3-position and a carbonyl group at the 4-position. We select merafloxacin (MERA) together with two structurally related analogs, denoted A1 and A2, which differ in their C7 substituents, following the nomenclature of the original experimental study by Sun et al~\cite{sun2021restriction}. MERA was previously identified as an inhibitor of $-1$ PRF in coronaviruses and was shown to restrict viral replication in cell-based assays, making it a relevant reference compound for RNA-targeted antiviral design~\cite{sun2021restriction,munshi2022identifying,yang2023discovery}. Compounds A1 and A2 are included as analogs with weak $-1$ PRF inhibition to compare how changes in ligand structure influence RNA binding and their effects on the RT and RU structures.

\begin{figure}
  \includegraphics[width=0.8\textwidth]{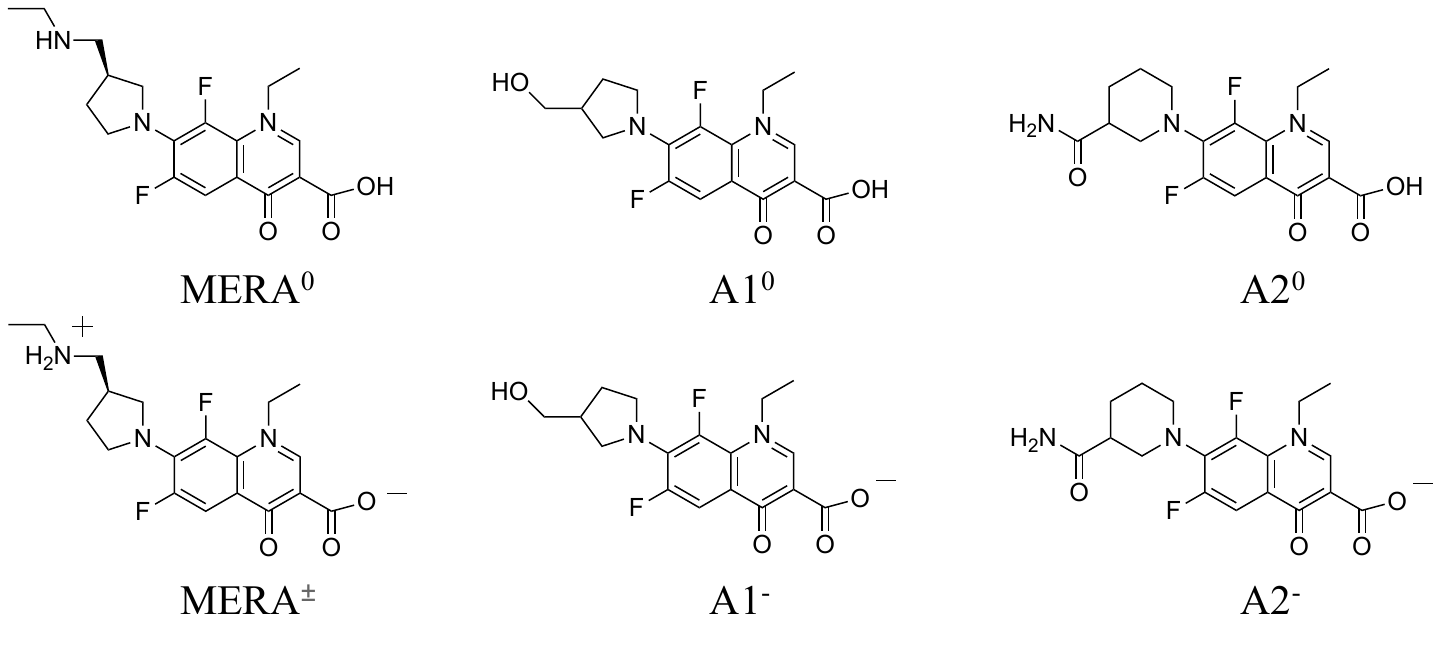}
  \caption{
    Ligands considered in this study. (left) Merafloxacin in its neutral (MERA$^{0}$) and zwitterionic (MERA$^{\pm}$) forms; (center) Analog 1 molecule in its neutral (A1$^{0}$) and anionic (A1$^{-}$) forms; (right) Analog 2 molecule in its neutral (A2$^{0}$) and anionic (A2$^{-}$) forms.
  }
  \label{fig:ligands}
\end{figure}

To account for physiological conditions, we consider the relevant protonation states of each ligand at a pH$\approx7$. Based on the reported dissociation constants of fluoroquinolones~\cite{van2014fluoroquinolone}, MERA is modeled in both its neutral (MERA$^{0}$) and zwitterionic (MERA$^{\pm}$) forms. In MERA$^{\pm}$, the quinolone carboxyl group is deprotonated, while the side-chain amine is protonated. By contrast, A1 and A2 were modeled in their neutral (A1$^{0}$, A2$^{0}$) and anionic (A1$^{-}$, A2$^{-}$) forms, consistent with the presence of a quinolone carboxyl group and the absence of a side-chain amine group (\rtab{tab:summary}). The three selected ligands in their neutral and ionized forms are shown in \rfig{fig:ligands}.

The RNA--ligand complexes are prepared using docking with AutoDock Vina~\cite{trott2010autodock,eberhardt2021autodock} and \texttt{vinardo}~\cite{quiroga2016vinardo} scoring function. For each ligand, docking calculations are carried out against both the RT and RU models of the pseudoknot, and the three top-ranked poses are retained as starting structures for MD simulations.

\begin{table}{%
  \scriptsize
  \begin{tabular}{lll}
    Abbreviation & RNA conformation & Ligand \\
    \hline
    RT              & Threaded   & ---                        \\
    RT-MERA$^{0}$   & Threaded   & Merafloxacin               \\
    RT-MERA$^{\pm}$ & Threaded   & Merafloxacin (zwitterion)  \\
    RT-A1$^{0}$      & Threaded   & Analog 1                  \\
    RT-A1$^{-}$      & Threaded   & Analog 1 (anion)        \\
    RT-A2$^{0}$      & Threaded   & Analog 2                  \\
    RT-A2$^{-}$      & Threaded   & Analog 2 (anion)        \\
    RU              & Unthreaded & ---                        \\
    RU-MERA$^{0}$   & Unthreaded & Merafloxacin               \\
    RU-MERA$^{\pm}$ & Unthreaded & Merafloxacin (zwitterion)  \\
    RU-A1$^{0}$      & Unthreaded & Analog 1                  \\
    RU-A1$^{-}$      & Unthreaded & Analog 1 (anion)        \\
    RU-A2$^{0}$      & Unthreaded & Analog 2                  \\
    RT-A2$^{-}$      & Unthreaded & Analog 2 (anion)        \\
  \end{tabular}
  }
  \caption{%
    The complete list of the systems considered in this work.
  }
  \label{tab:summary}
\end{table}

\subsection{Molecular Dynamics}
We perform MD simulations for RT and RU without any ligands, as well as for each RNA--ligand complex with ligands in neutral or ionized states (\rtab{tab:summary}). The simulations are carried out in GROMACS 2024.2 using the CHARMM36 force field~\cite{denning2011impact} (July 2022 release) together with the CHARMM-modified TIP3P water model. Ligand topologies and parameters are assigned using the CHARMM General Force Field (CGenFF).~\cite{vanommeslaeghe2010charmm} After system preparation, each system undergoes 1 ns of NVT equilibration followed by 1 ns of NPT equilibration. Temperature is controlled at 310 K with the velocity-rescaling thermostat~\cite{bussi2007canonical}, and pressure is maintained at 1 atm with stochastic cell rescaling~\cite{bernetti2020pressure}. Unrestrained production runs are then performed for 0.4--3 $\mu$s with a 2 fs time step. We use a dodecahedron box with a 1.5 nm margin. The resulting trajectories were analyzed using the MDTraj library~\cite{McGibbon2015MDTraj}.

To assess the structural stability and conformational dynamics of the pseudoknot in the MD simulations, we compute the root-mean-square deviation (RMSD) of the full RNA structure, per-residue root-mean-square fluctuation (RMSF), and the fraction of native contacts, $Q$, for each of the three RNA stems over time. These quantities are calculated from mass-centered trajectories after structural alignment.

$Q$ is evaluated following the definition of Best, Hummer, and Eaton~\cite{best2013native}. For a given conformation $\bx$, $Q(\bx)$ is computed as a smooth contact-based measure relative to the reference native structure, using all pairs of heavy atoms $(k,l)$ belonging to stem residues separated by more than three positions in sequence, and whose reference distance satisfied $x_{kl}^0 < 0.45$ nm. The contact contribution of each pair is weighted according to:
\begin{equation}
  Q(\bx)=\frac{1}{c}\sum_{k>l}^n \ppar*{1+\e^{\alpha(x_{kl}-\gamma
  x_{kl}^0)}}^{-1},
\end{equation}
where $c$ is the number of native contacts, $x_{kl}$ is the heavy-atom pairwise distance between atoms $k$ and $l$, $x_{kl}^0$ is the distance in the reference structure, and the parameters are set to $\alpha = 50$ nm$^{-1}$ and $\gamma = 1.5$, as recommended for all-atom simulations~\cite{best2013native}. $Q$ is calculated only for the stem regions of the pseudoknot in order to specifically monitor the preservation of the native helical core.

\subsection{Machine Learning}
We define the input descriptors for SM as pairwise distances between 1-position nitrogen atoms (N1) of the nucleobases. The N1 atoms occupy conserved positions within the Watson-Crick edge of the bases: in pyrimidines (C, U), N1 forms the glycosidic linkage to the ribose, whereas in purines (A, G), N1 directly participates in canonical hydrogen bonding. As a result, N1-N1 distances provide a consistent and physically interpretable measure of base pairing, stacking rearrangements, and tertiary contact formation, and thus encode the information about conformational transitions during RNA dynamics~\cite{bottaro2014role}. The descriptors are extracted from the MD trajectories every 50 ps (i.e., every fifth frame from trajectories saved at 10 ps intervals). To reduce contributions from thermal noise, we apply variance-based preprocessing and, for each system, retain the 100 descriptors with the highest variance. This procedure preserves information about relevant conformational changes and maintains a fixed NN input size across our systems.

We employ NNs with two hidden layers of 75 nodes each and ELU activation
functions. Each network contains approximately 13,000 trainable parameters. The
output comprises two CVs. Training uses batches of 1000 samples, with 20\% of
the data reserved for validation. Training proceeds until early stopping is
triggered when the validation loss reaches $10^{-4}$ precision with a patience
of 20 epochs. We use the Adam optimizer with a learning rate of $10^{-3}$ and
default parameters. In all cases, we enforce orthogonality on NN parameters.

For the spectral loss (\req{eq:loss-eig}), we use $k=3$ dominant eigenvalues
calculated from the symmetric conjugate of the Markov transition matrix
(\req{eq:markov}). A detailed discussion is given in Sec.~S1 in SI. 
Following standard practice in diffusion map methods, we set
the Gaussian kernel bandwidth $\varepsilon$ to the median of pairwise Euclidean
distances.

To assess the quality of the training protocol, we analyze the stability of
the loss function across epochs. All systems converge within $\sim$100 epochs
with agreement between the training and validation losses, ruling out
overfitting (\rmfig{1}). The spectral loss $L_\lambda$ (negative cumulative spectral
weight) dominates the total loss, while the decorrelation ($L_{\mathrm{D}}$)
and Laplacian ($L_{\mathrm{L}}$) losses remain near zero throughout, indicating
that the two learned CVs are easily kept decorrelated and smooth (\rmfig{2} and \rmfig{3}).
Additionally, the Gaussian kernel bandwidth $\varepsilon$, estimated based on the median of
pairwise distances between samples in CV space (\req{eq:kernel}), converges to a narrow range
(0.15--0.25) for all systems (\rmfig{4}), confirming that the orthogonal weight constraints in
the encoder prevent the scale collapse and, thus, the degeneration of the spectral
characteristics of the learned CVs.

We use SM implemented in our development version of the \texttt{spectre} package~\cite{spectre},
programmed using the PyTorch~\cite{paszke2019pytorch} and Lightning frameworks~\cite{lightning}.

\subsection{Free-Energy Landscapes}
\label{subsec:fel}
We compute free-energy landscapes as $F(\bz) = -\beta^{-1}\log P(\bz)$, where
$\beta=1/\kT$ is the inverse temperature with the Boltzmann constant denoted as
$\kb$ and temperature as $T$. We calculate the marginal probability density $P(\bz)$ in
CV space by kernel density estimation (KDE). In KDE, we use Gaussian kernels
with bandwidth estimated from Scott's rule. 

We estimate free-energy differences as $\Delta F_{\mathrm{A \to B}} = -\beta^{-1} \log \ppar*{\int_\mathrm{A} \dz \e^{-\beta F(\bz)} / \int_\mathrm{B} \dz \e^{-\beta F(\bz)}}$, where the integrals go over states A and B. Integration regions over states are computed on a grid by finding local minima and assigning each grid point to the nearest minimum. We select only grid points below 20 kJ/mol. Then, $\Delta F_{A \to B}$ is calculated by iterating over each pair of states. As grid points are weighted by the Boltzmann factor, the main contribution to the integrals comes from regions near the minima, not at the separation.

For analysis, we identify MFEPs between each
minimum by performing path optimization on a FEL estimated
via KDE. In contrast to standard string optimization
algorithms~\cite{maragliano2006string,maragliano2006temperature}, rather than
computing forces on-the-fly from MD, we keep the FEL fixed.
The force on each interior path point combines perpendicular gradient descent
and spring forces $\bz_i^{(t+1)} = \bz_i^{(t)} + \alpha \ppar*{ -\nabla_\perp
F(\bz_i^{(t)}) + \kappa F_s }$, where $\bz_i^{(t)}$ is the position of the
$i$-th path point at iteration $t$. The force combines two terms: the perpendicular component of the free-energy gradient $\nabla_\perp
F(\bz_i)$ and the
spring force $F_s = \bz_{i+1}^{(t)} - 2\bz_i^{(t)} + \bz_{i-1}^{(t)}$ with
spring constant $\kappa=0.1$ that maintains equal spacing between adjacent
points. We use a learning rate $\alpha=0.05$. Around each path point, we select
10 closest samples to determine structural changes along each MFEP. We then
estimate barriers between states and the related errors. For a detailed description,
see Sec.~S4 in SI.

\section{Results and Discussion}
\subsection{Dynamics of the Ligand-Free Pseudoknot Topologies}
\begin{figure}[h]
  \includegraphics[width=0.9\textwidth]{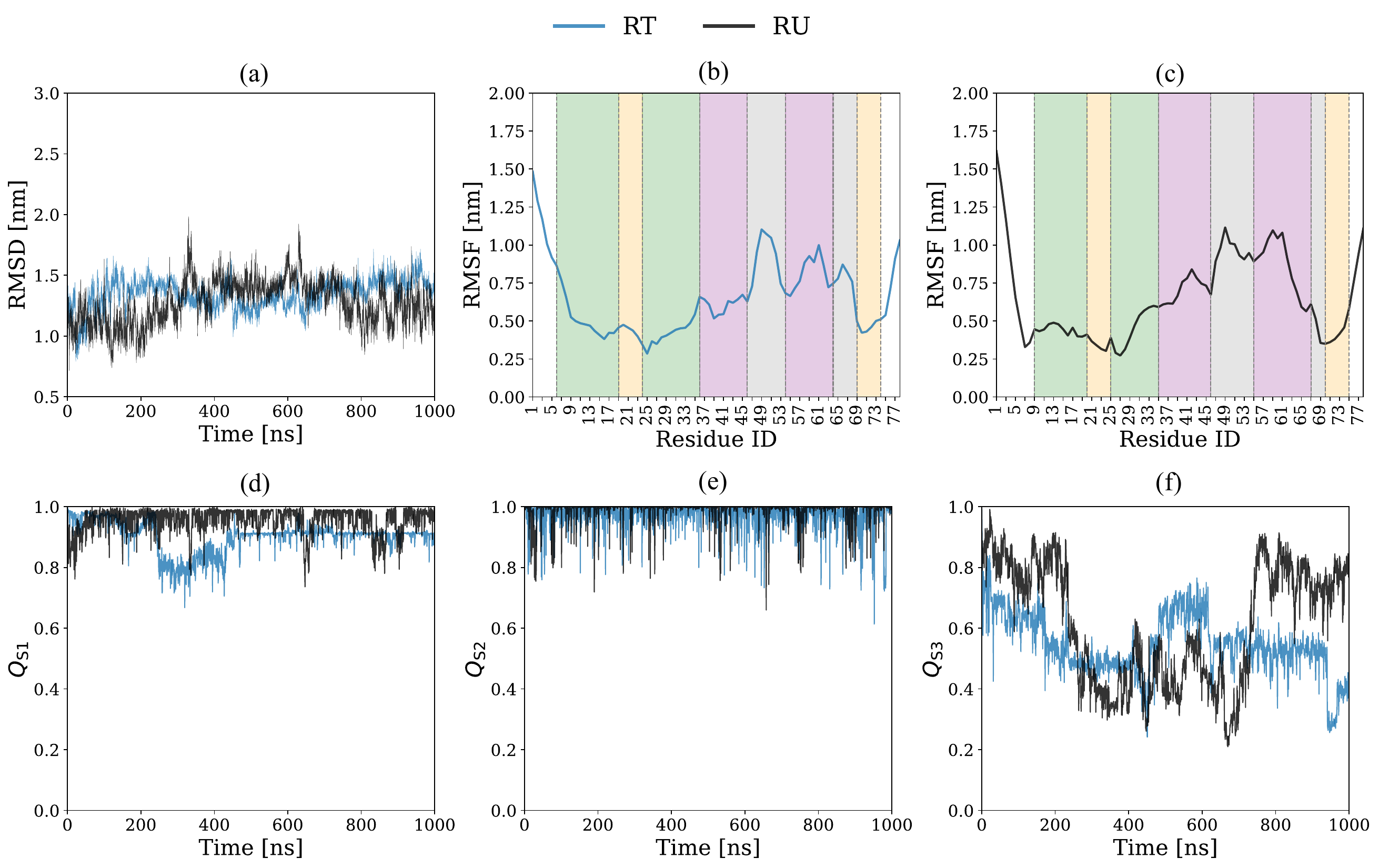}
  \caption{%
    Comparison of structural stability and conformational dynamics of the
    ligand-free threaded (RT) and unthreaded (RU) SARS-CoV-2 pseudoknot models.
    (a) RMSD of the overall pseudoknot structure. (b,c) RMSF values for RT and RU,
    respectively. Colors denote the pseudoknot regions: 5$'$- and 3$'$-end tails
    (white), S1 (green), S2 (orange), S3 (purple), and L1--L2 (gray). (d--f)
    Fraction of native contacts, $Q$, calculated for S1, S2, and S3,
    respectively.
  }
  \label{fig:md-stat-rt-ru}
\end{figure}

Before analyzing the ligand-containing trajectories, we first characterize the dynamics of ligand-free RT and RU pseudoknots. During a 1$\mu$s-long MD simulation, both RT and RU remain globally folded (\rsfig{fig:md-stat-rt-ru}{a}). Despite the topological difference associated with 5$'$-end threading, the two models exhibit a similar pattern of residue fluctuations. In particular, the largest fluctuations occur in the S3 (purple) and L2 (gray) regions, whereas the S1 (green) and S2 (orange) regions remain relatively stable (\rsfig{fig:md-stat-rt-ru}{b,c}), consistent with previous studies~\cite{yan2022length}. The high flexibility of S3 is further supported by the fraction of native contacts, where $Q_{\mathrm{S3}}$ decreases to $\sim$0.2 in both RT and RU (\rsfig{fig:md-stat-rt-ru}{f}), while $Q_{\mathrm{S1}}$ and $Q_{\mathrm{S2}}$ remain closer to 1 (\rsfig{fig:md-stat-rt-ru}{d,e}). Below, we treat these ligand-free trajectories as the reference against which ligand-induced perturbations were evaluated.

\subsection{Ligand-Induced Destabilization Is Topology-Selective}
To assess the influence of ligand binding on pseudoknot stability, we performed MD simulations of duration 0.4--1~$\mu$s for RT and RU forms while adding each ligand shown in \rfig{fig:ligands}. For each RNA--ligand pair, three separate MD simulations were initiated from the three top-ranked docking poses shown in \rmfig{5} to assess the sensitivity of the observed RNA dynamics to the initial ligand placement and to test the reproducibility of ligand-induced structural responses. Across the full dataset of trajectories (\rmfig{6}--\rmfig{9}), in most cases RNA remain structurally similar to the corresponding ligand-free trajectories, exhibiting only modest changes. However, a subset of systems demonstrated clear ligand-induced distortions of the pseudoknot conformation. The most pronounced effects were observed for RT-MERA$^{0}$ (top-1 pose), RT-MERA$^{\pm}$ (top-1 pose), RT-A1$^{0}$ (top-1 pose), RU-MERA$^{\pm}$ (top-2 pose), and RU-A1$^{0}$ (top-1 pose); the pose number in parentheses identifies the corresponding docking-derived initial ligand position. To further characterize their long-timescale behavior, these simulations were extended to 3~$\mu$s for RT-MERA$^{0}$, RT-MERA$^{\pm}$, and RT-A1$^{0}$, and to 2~$\mu$s for RU-MERA$^{\pm}$ and RU-A1$^{0}$. We analyze these five RNA--ligand systems below.

\begin{figure}[h]
  \includegraphics[width=0.9\textwidth]{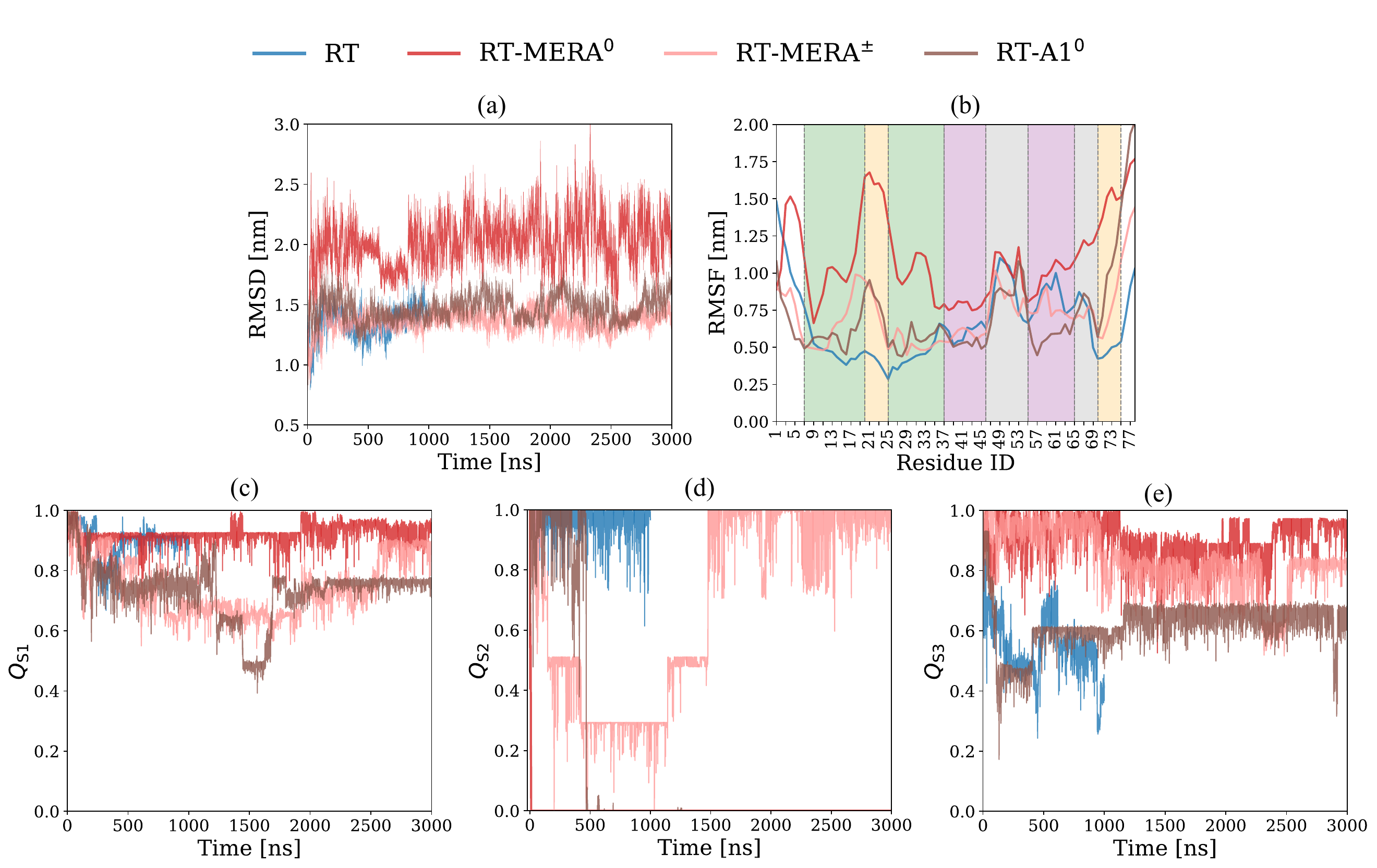}
  \caption{
    Structural stability and conformational dynamics of the threaded (RT)
    topology of the SARS-CoV-2 pseudoknot for the RNA--ligand systems showing the
    largest conformational distortions. (a) RMSD of the overall pseudoknot structure
    for RT--ligands. (b) RMSF values for the RT--ligand trajectories. Colors
    denote the pseudoknot regions: 5$'$- and 3$'$-end tails (white), S1 (green), S2
    (orange), S3 (purple), and L1--L2 (gray). (c--e) $Q$, calculated for S1,
    S2, and S3, respectively.
  }
  \label{fig:rt-stat-disc}
\end{figure}

Among all systems, MERA$^{0}$ produces the most pronounced structural perturbation (\rsfig{fig:rt-stat-disc}{a}). This effect is localized primarily in the S2 region, which becomes markedly more flexible when the ligand is present (\rsfig{fig:rt-stat-disc}{b}). MERA$^{\pm}$ and A1$^{0}$ also enhance fluctuations in S2, although the effect is more moderate. In agreement with these observations, $Q_{\mathrm{S2}}$ drops significantly in all RT--ligand trajectories (\rsfig{fig:rt-stat-disc}{d}), whereas $Q_{\mathrm{S1}}$ and $Q_{\mathrm{S3}}$ (\rsfig{fig:rt-stat-disc}{c,e}) remain higher and do not fall to zero. Importantly, the S2 stem is partially restored after $\approx 1500$ ns in the RT-MERA$^{\pm}$ simulation, while in the RT-MERA$^{0}$ and RT-A1$^{0}$ simulations it drops to zero quickly and remains disrupted for the rest of time. These results suggest that ligand binding may selectively weaken a structurally vulnerable S2 region of the threaded pseudoknot, thereby causing breakage of the secondary structure. Moreover, the mechanism appears to be allosteric: upon S2 breakage, the ligand is bound to a different stem, as illustrated in \rmfig{10}{a--c}.

\begin{figure}[h]
  \includegraphics[width=0.9\textwidth]{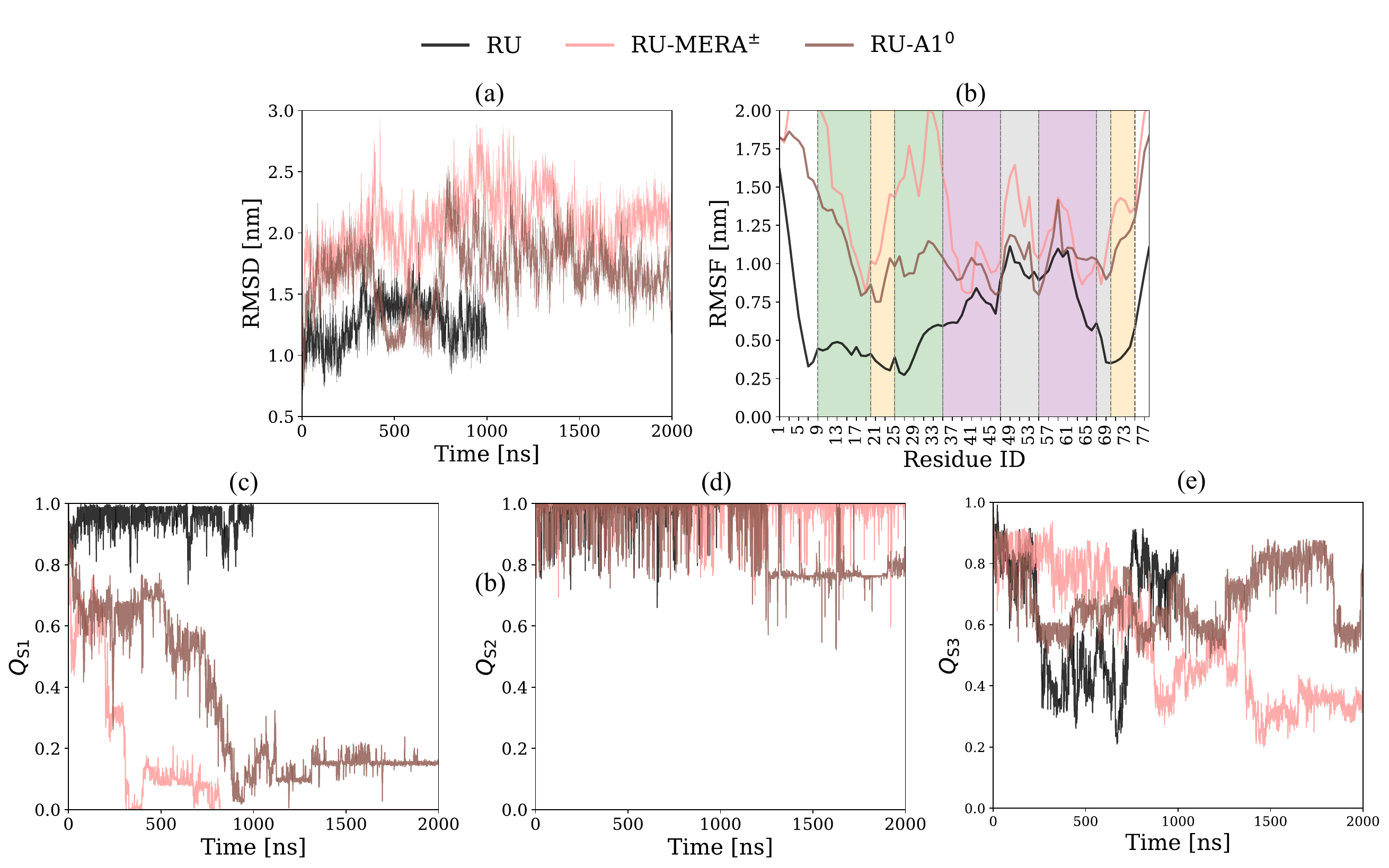}
  \caption{
    Structural stability and conformational dynamics of the unthreaded (RU)
    topology of the SARS-CoV-2 pseudoknot for the RNA--ligand systems showing the
    largest conformational distortions. (a) RMSD of the overall pseudoknot
    structure for RU--ligands. (b) Per-residue RMSF profile for the RU--ligand
    trajectories. Colors denote the pseudoknot regions: 5$'$- and 3$'$-end tails
    (white), S1 (green), S2 (orange), S3 (purple), and L1--L2 (gray). (c--e)
    $Q$, calculated separately for S1, S2, and S3, respectively.
  }
  \label{fig:ru-stat-disc}
\end{figure}

For the unthreaded RNA--ligand systems, both RU-MERA$^{\pm}$ and RU-A1$^{0}$ exhibit RMSD deviations over the course of the MD simulations (\rsfig{fig:ru-stat-disc}{a}). Unlike the RT--ligand systems, the RU--ligand systems display substantial distortions primarily in the S1 and S3 regions (\rsfig{fig:ru-stat-disc}{b}). This behavior is consistent with the $Q$ values. In the RU-MERA$^{\pm}$ trajectory, $Q_{\mathrm{S1}}$ eventually decreases to zero, and the native contacts of S1 do not recover. Similarly, in RU-A1$^{0}$, $Q_{\mathrm{S1}}$ also decreases (\rsfig{fig:ru-stat-disc}{c}). The events of S1 breakage for these two systems are illustrated in \rmfig{10}{d,e}. S2 remains comparatively stable in all RU-containing systems, with $Q_{\mathrm{S2}}$ staying close to that of the ligand-free RU reference (\rsfig{fig:ru-stat-disc}{d}). S3 remains the most flexible stem region, and $Q_{\mathrm{S3}}$ decreases markedly during the simulation, although it does not reach zero (\rsfig{fig:ru-stat-disc}{e}).

RMSD, RMSF, and $Q$ values indicate that substantial pseudoknot reorganization occurs in only a minority of ligand-containing MD trajectories. Thus, pseudoknot destabilization is not a generic consequence of ligand proximity. In fact, ligands frequently migrate over long distances during simulations because of the high flexibility of RNA and its shallow binding pockets (see \rmfig{11}). Instead, the specific combination of ligand position, type, protonation state, and pseudoknot topology is important to induce conformational changes. The non-perturbed RNA trajectories, therefore, serve as an important baseline, indicating that the pronounced conformational changes observed in the representative systems reflect selective RNA--ligand recognition events. 

Below, we apply SM to further elucidate the pseudoknot conformational dynamics in the selected MD trajectories (\rfig{fig:rt-stat-disc} and \rfig{fig:ru-stat-disc}), and compare them with the ligand-free RT and RU reference trajectories (\rfig{fig:md-stat-rt-ru}). For contrast, we also include the RT-A2$^0$ (top pose-1) trajectory, which did not show a significant structural perturbation.

\begin{figure}[p]
  \includegraphics{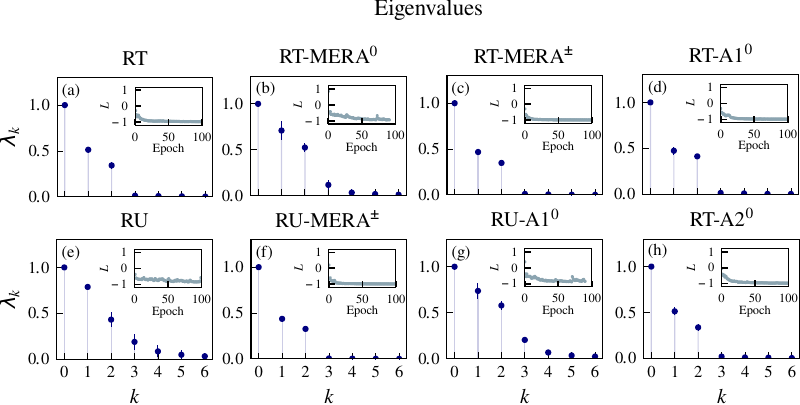}
  \caption{%
    Eigenvalues $\lambda_k$ corresponding to the converged CVs after SM
    learning. Dots and error bars show means and standard deviations,
    respectively, calculated over the last 50 epochs. Insets show the training
    loss $L$ during 100 epochs.
    }
  \label{fig:eigval}
\end{figure}
\begin{figure}[p]
  \includegraphics{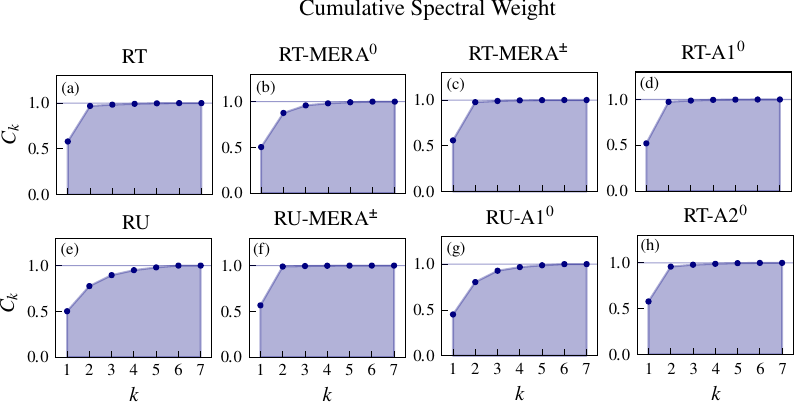}
  \caption{%
    Cumulative spectral weights of the dominant eigenvalues ($\pset{\lambda_i}$
    for $i>0$), $c_k = \sum_{i=1 \le k} \lambda_i / \sum_{i=1} \lambda_i$,
    after training.
    }
  \label{fig:eigval-ck}
\end{figure}

\subsection{Slow Modes Distinguish Threaded and Unthreaded Topologies}
\label{sec:spectra}
The eigenspectra of the converged CVs reveal a fundamental dynamical distinction between the threaded and unthreaded pseudoknot topologies (\rfig{fig:eigval}). Apart from RT-MERA$^0$ (\rsfig{fig:eigval}{b}), the RT systems show clear spectral gaps ($\Delta \lambda = \lambda_k - \lambda_{k-1}$), having the dominant eigenvalues $\lambda_1$ and $\lambda_2 \approx 0.5$, and demonstrate the fast saturation of the cumulative spectral weights (\rsfig{fig:eigval-ck}). After that, the remaining eigenvalues drop rapidly to 0. This indicates three well-separated slow modes and, correspondingly, three dominant metastable states: S$_0$, S$_1$, and S$_2$. For the RT-MERA$^0$ system (\rsfig{fig:eigval}{b}), the spectral gap is slightly smaller, suggesting the lack of considerable timescale separation between states. In contrast, the ligand-free RU shows a gradual eigenvalue decay with no clear spectral gap (\rsfig{fig:eigval}{e}), and the cumulative spectral weight saturates slowly (\rsfig{fig:eigval-ck}{e}). This indicates that, for RU, there are no distinct slow modes; instead, all conformational transitions are relatively fast, as the system is essentially a broad superstate composed of smaller states separated by negligible free-energy barriers. Interestingly, introducing MERA$^\pm$ and A1$^0$ ligands to RU recovers spectral gaps resembling those of the RT systems, suggesting that the inhibitors have a strong effect on the dynamics of the unthreaded system. This effect is especially noticeable for the MERA zwitterion (\rsfig{fig:eigval}{f}), demonstrating that MERA$^{\pm}$ slows the dynamical behavior of the unthreaded pseudoknot topology.

\begin{figure}[p]
  \includegraphics[width=\textwidth]{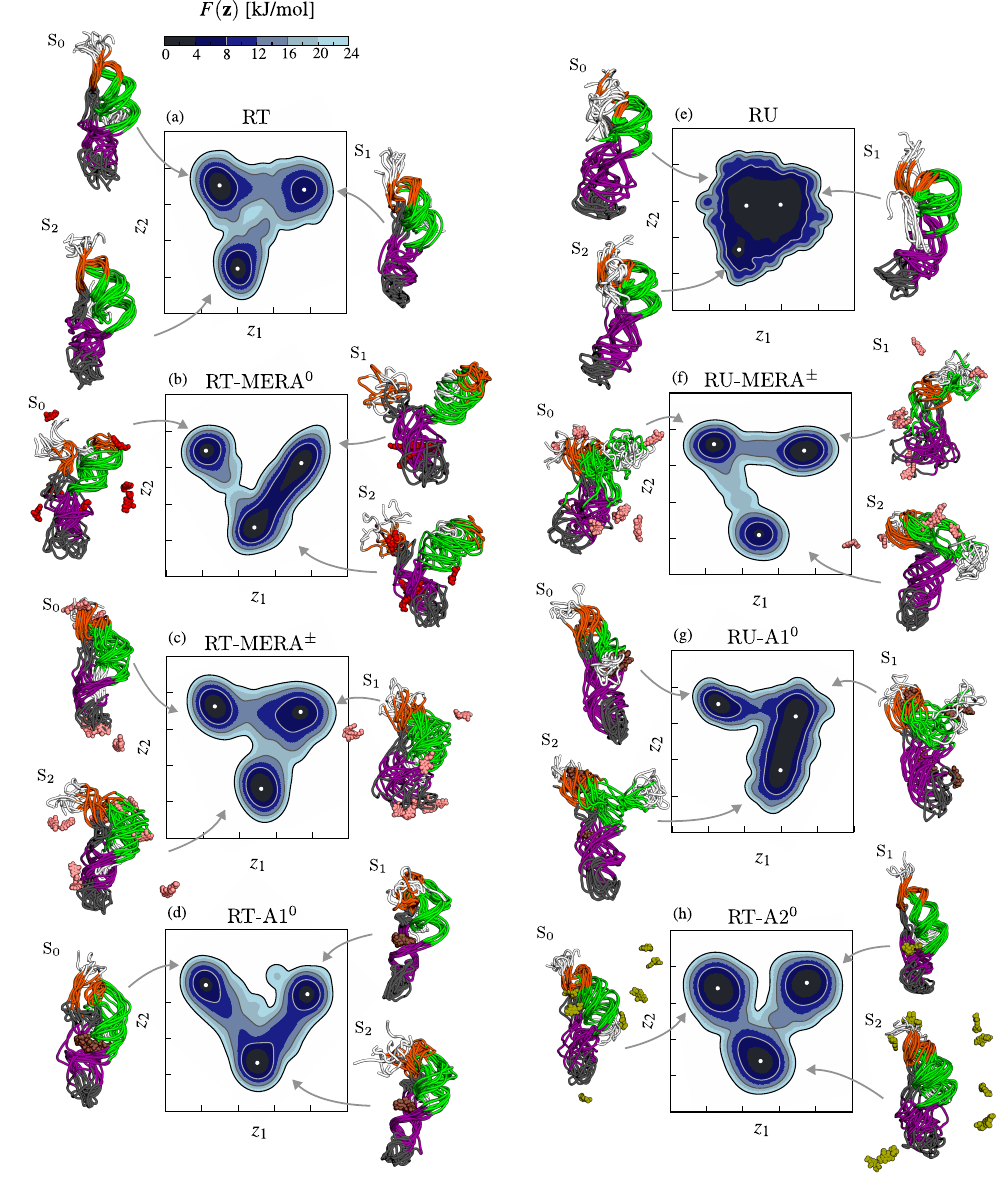}
  \caption{%
    Free-energy landscapes $F(\bz)$ reconstructed in CV space $\bz=(z_1,z_2)$
    learned by SM for each system. Each state is labeled by S$_0$, S$_1$, and
    S$_2$. The rotation of each landscape is set so that the state labels are
    ordered clockwise. The ensembles corresponding to the metastable states
    depict 10 structures closest to the minima (white dots).
  }
  \label{fig:cycle}
\end{figure}

\subsection{Ligands Reshape the Pseudoknot Free-Energy Landscape}
Based on FELs constructed from the learned CVs, we find that the ligands can significantly reshape the conformational dynamics of the pseudoknot (\rfig{fig:cycle}). In the ligand-free RT, we observe three metastable basins in a symmetric triangular arrangement, with energy barriers between them of around $\sim$7--14 kJ/mol (\rsfig{fig:cycle}{a}). Additionally, the FEL allows this system to transition between all states without a discernible bottleneck region. By contrast, adding MERA$^0$ to the threaded form significantly reshapes its FEL (\rsfig{fig:cycle}{b}). RT-MERA$^0$ is characterized by an asymmetric landscape with one significantly lowered barrier ($F^\dagger_{\mathrm{S}_1 \to \mathrm{S}_2} \approx 3$ kJ/mol, see \rtab{tab:energy}). Both S$_1$ and S$_2$ states, which correspond to this fast transition, are the most disrupted conformations relative to ligand-free RT (\rsfig{fig:cycle}{b}), with S2 fluctuating the most. In both states, the ligand is bound to the S3 stem. This can be linked to the strong allosteric S2 stem disruption upon ligand binding and is in agreement with RMSF and $Q_{S2}$ plots (see \rfig{fig:rt-stat-disc}{b,d}). Although the FELs of the RT-MERA$^\pm$ (\rsfig{fig:cycle}{c}) and RT-A1$^0$ (\rsfig{fig:cycle}{d}) systems have larger barrier $F^\dagger_{\mathrm{S}_1 \to \mathrm{S}_2} > 7$ kJ/mol (see \rtab{tab:energy}), they exhibit asymmetry similar to RT-MERA$^0$ and lack direct transitions between two of the three states. Notably, the A1$^0$ ligand did not move far from its initial position during the full MD simulation, revealing a relatively stable binding pocket inside the S3 strands.

In the ligand-free RU (\rsfig{fig:cycle}{e}), the system is located in a flat FEL, with all barriers $F^\dagger < 4$ kJ/mol (see \rtab{tab:energy}). As mentioned in \rsct{sec:spectra}, MERA$^{\pm}$ significantly affects the spectral gap and transforms the RU landscape into three well-separated basins with substantial barriers (11--18 kJ/mol, see \rtab{tab:energy}), thereby slowing conformational dynamics (\rsfig{fig:cycle}{f}). The FEL of RU-A1$^0$ (\rsfig{fig:cycle}{g}) also features three basins, but similar to RT-MERA$^0$, the barrier of $F^\dagger_{\mathrm{S}_1 \to \mathrm{S}_2} < 1$ kJ/mol (see \rtab{tab:energy}) is very low. In both complexes, the states mainly differ in the S1 and S3 stem regions, in agreement with the corresponding RMSF, $Q_{S1}$, and $Q_{S3}$ plots (see \rsfig{fig:ru-stat-disc}{b,d}).

As an example of a relatively undistorted RNA--ligand system, we show the FEL of RT-A2$^0$ (\rsfig{fig:cycle}{h}). Most ligand conformations in RT-A2$^0$ diffuse around the pseudoknot rather than bind at a specific location. Compared with ligand-free RT, the RNA states and FEL shape here are qualitatively similar, underscoring that A2$^0$ has the weakest effect on RT.

The FELs analyzed here are constructed from the subset of MD trajectories that sampled substantial RNA distortion, and thus they characterize the conformational space each ligand makes accessible rather than a globally converged RNA—ligand landscape. On this basis, the SM analysis provides a realistic picture of pseudoknot dynamics, even for highly flexible RNA. The ligand-induced perturbations we identify correlate with the experimentally measured $-1$ PRF inhibition efficacy, that follows the order MERA $>$ A1 $>>$ A2.~\cite{sun2021restriction} MERA induces conformational rearrangements of the pseudoknot more frequently than A1, and both MERA and A1 disrupt native contacts within the RNA stems and stabilize metastable states absent from the ligand-free RNA, whereas A2 does not. Together, these results make an important contribution to a design principles for antiviral inhibitors: the most effective ligands are those that change the RNA FEL most strongly, destabilizing the pseudoknot conformation required for efficient $-1$ PRF.

\begin{table}[t]
\scriptsize\centering
  \caption{%
    Absolute free-energy differences ($|\Delta F|$) and free-energy barriers ($F^\dagger$)
    between the identified states. Errors are denotes by $\sigma$. Missing barrier values (marked by
    ``---'') indicate missing direct path and a transition
    through an intermediate as obtained by estimating MFEPs (see \rfig{fig:cycle}). See Sec.~S4 in SI for
    more details.}
\vspace{0.2cm}
\begin{tabular}{l c c c c c c c}
\hline\\
System & Transition 
  & $|\Delta F|$ [kJ/mol] 
  & $F^\dagger$ [kJ/mol]
  & $\sigma$ [kJ/mol]
  & $F^\dagger$ [kJ/mol]
  & $\sigma$ [kJ/mol] \\
& & & \multicolumn{2}{c}{$A\to B$} & \multicolumn{2}{c}{$B\to A$} \\[0.2cm]
\hline
                & $\mathrm{S}_0 \leftrightarrow \mathrm{S}_1$  & 4.7 & 12.4 & 0.8 & 7.7  & 0.8\\
RT              & $\mathrm{S}_1 \leftrightarrow \mathrm{S}_2$  & 2.2 & 14.0 & 1.3 & 12.2 & 1.3\\
                & $\mathrm{S}_0 \leftrightarrow \mathrm{S}_2$  & 2.2 & 13.1 & 1.5 & 10.2 & 1.5\\
\hline
                & $\mathrm{S}_0 \leftrightarrow \mathrm{S}_1$  & 1.1 & ---  & --- & --- & --- \\
RT-MERA$^0$     & $\mathrm{S}_1 \leftrightarrow \mathrm{S}_2$  & 0.9 & 3.9  & 0.1 & 3.1 & 0.1 \\
                & $\mathrm{S}_0 \leftrightarrow \mathrm{S}_2$  & 4.5 & 18.6 & 1.4 & 16.1 & 1.5 \\
\hline
                & $\mathrm{S}_0 \leftrightarrow \mathrm{S}_1$  & 1.8 & 11.7 & 0.5 & 9.8  & 0.5 \\
RT-MERA$^\pm$   & $\mathrm{S}_1 \leftrightarrow \mathrm{S}_2$  & 0.5 & 15.6 & 1.5 & 12.9 & 1.4 \\
                & $\mathrm{S}_0 \leftrightarrow \mathrm{S}_2$  & 0.4 & ---  & --- & --- & --- \\
\hline
                & $\mathrm{S}_0 \leftrightarrow \mathrm{S}_1$  & 0.8  & ---  & --- & --- & --- \\
RT-A1$^0$       & $\mathrm{S}_1 \leftrightarrow \mathrm{S}_2$  & 2.7 & 9.6  & 0.2 & 7.5  & 0.2 \\
                & $\mathrm{S}_0 \leftrightarrow \mathrm{S}_2$  & 3.2 & 11.9 & 0.4 & 9.2  & 0.4 \\
\hline
                & $\mathrm{S}_0 \leftrightarrow \mathrm{S}_1$  & 0.1 & 0.5  & 0.3 & 0.4  & 0.3 \\
RU              & $\mathrm{S}_1 \leftrightarrow \mathrm{S}_2$  & 3.5 & 3.6  & 0.4 & 0.5  & 0.5 \\
                & $\mathrm{S}_0 \leftrightarrow \mathrm{S}_2$  & 4.5 & 4.3  & 0.5 & 0.5  & 0.6 \\
\hline
                & $\mathrm{S}_0 \leftrightarrow \mathrm{S}_1$  & 0.3 & 17.8 & 1.2 & 14.4 & 1.2  \\
RU-MERA$^\pm$   & $\mathrm{S}_1 \leftrightarrow \mathrm{S}_2$  & 2.8 & ---  & --- & --- & ---  \\
                & $\mathrm{S}_0 \leftrightarrow \mathrm{S}_2$  & 2.0 & 13.1 & 0.7 & 11.0 & 0.8 \\
\hline
                & $\mathrm{S}_0 \leftrightarrow \mathrm{S}_1$  & 2.0 & 10.5 & 1.0 & 11.2 & 0.9  \\
RU-A1$^0$       & $\mathrm{S}_1 \leftrightarrow \mathrm{S}_2$  & 1.1 & 0.9  & 0.1 & 0.1  & 0.1 \\
                & $\mathrm{S}_0 \leftrightarrow \mathrm{S}_2$  & 0.4 & ---  & --- & --- & ---  \\
\hline
                & $\mathrm{S}_0 \leftrightarrow \mathrm{S}_1$  & 0.8 & --- & --- & --- & ---   \\
RT-A2$^0$       & $\mathrm{S}_1 \leftrightarrow \mathrm{S}_2$  & 1.7 & 15.7 & 1.5 & 14.4 & 1.5 \\
                & $\mathrm{S}_0 \leftrightarrow \mathrm{S}_2$  & 2.1 & 13.3 & 0.9 & 12.2 & 0.9 \\
\hline
  \end{tabular}
  \label{tab:energy}
\end{table}

\subsection{Merafloxacin Protonation Affects the Binding Mechanism}
An interesting result is that significant differences exist between the FELs of MERA$^0$ and MERA$^{\pm}$ interacting with the pseudoknot (\rsfig{fig:cycle}{b,c}). These two protonation states of the same inhibitor produce qualitatively different FELs for the same RNA topology. In the threaded form, MERA$^0$ creates an asymmetric landscape with a very low $F^\dagger_{\mathrm{S}_1 \to \mathrm{S}_2}$ barrier, whereas MERA$^{\pm}$ produces a more symmetric three-basin landscape with a significantly higher $F^\dagger_{\mathrm{S}_1 \to \mathrm{S}_2}$ barrier. For the unthreaded form, only the zwitterionic MERA$^{\pm}$ significantly perturbs the FEL of RU. Thus, this confirms that the physiological protonation state of MERA is essential for accurate prediction of RNA--drug interactions.

\section{Conclusions}
We have used molecular dynamics (MD) simulations and spectral map (SM), a thermodynamics-driven ML method, to characterize the conformational dynamics of the SARS-CoV-2 RNA pseudoknot, regulating the $-1$ programmed ribosomal frameshifting ($-1$ PRF) mechanism in virus translation. We simulated two distinct RNA fold topologies in complex with merafloxacin and its two structural analogs, for two ligand protonation states. By learning CVs that directly target slow dynamic modes of the RNA--ligand systems, we obtained free-energy landscapes (FELs) that link the chemical structure of ligands, their protonated states, and RNA topology to the rare conformational transitions that underlie pseudoknot function.

We have shown that ligand-induced destabilization is topology-selective. The substantial decreases in the fraction of native stem contacts, $Q$, indicate that in the threaded pseudoknot, the strongest distortion occurs in the S2 stem, whereas in the unthreaded pseudoknot, the distortion is concentrated in the S1 and S3 stems. Moreover, for the threaded RNA, we observed that ligand-induced RNA distortion proceeds via an allosteric-like mechanism.

The degree to which each ligand alters the conformational dynamics of the pseudoknot, as illustrated by the FELs, matches the experimental ranking of $-1$ PRF inhibition efficacy reported by Sun et al~\cite{sun2021restriction}. In contrast, the neutral and zwitterionic forms of merafloxacin create qualitatively different FELs for the same RNA topology. This shows that the protonation state of the ligand should be considered in mechanistic models of RNA--ligand recognition.

These findings support a topology- and pose-dependent view of ligand recognition by the SARS-CoV-2 pseudoknot and, more broadly, by structured RNA. In contrast to proteins, which typically present well-defined binding pockets~\cite{rydzewski2017ligand,bernetti2019kinetics}, the RNA pseudoknot offers shallow, transient interaction sites: ligands placed in a docked pose often migrated during the simulation, and the same RNA--ligand pair did not always produce the same response across trajectories initiated from different poses. We do not interpret the perturbations we observe as a reproducible bound state but as transient binding poses that couple to structurally sensitive regions of the RNA. We view the simulations presented here as a mechanistic exploration of candidate microscopic states rather than a one-to-one structural assignment of experimentally observed RNA--ligand complexes.

From a methodological standpoint, our results demonstrate that SM can recover the slow modes governing the conformational dynamics of structured RNA. The resulting FELs provide a quantitative, mechanistic bridge between the chemical structure of the ligand and rare conformational transitions of RNA. SM can thus be used alongside the growing set of ML methods developed for RNA~\cite{sacco2026machine}.

\begin{acknowledgement}
M.I. acknowledges financial support from JSPS KAKENHI Grant Numbers JP24K23894 and JP22KJ2450, and computational resources provided by the Research Institute for Information Technology at Kyushu University, including access through the HPCI System Research Project (hp250160). M.I. also gratefully acknowledges Prof. Asako Murata for insightful discussions on the experimental aspects of RNA-targeted drug discovery and Prof. Yuriko Aoki for her mentorship and guidance in research.
J.R. acknowledges funding from the Ministry of Science and Higher Education in Poland and the National Science Center in Poland (Sonata 2021/43/D/ST4/00920, ``Statistical Learning of Slow Collective Variables from Atomistic Simulations'').

\end{acknowledgement}

\bibliography{main.bib}

\newpage

\section*{TOC}
\begin{figure}
    \centering
    \includegraphics{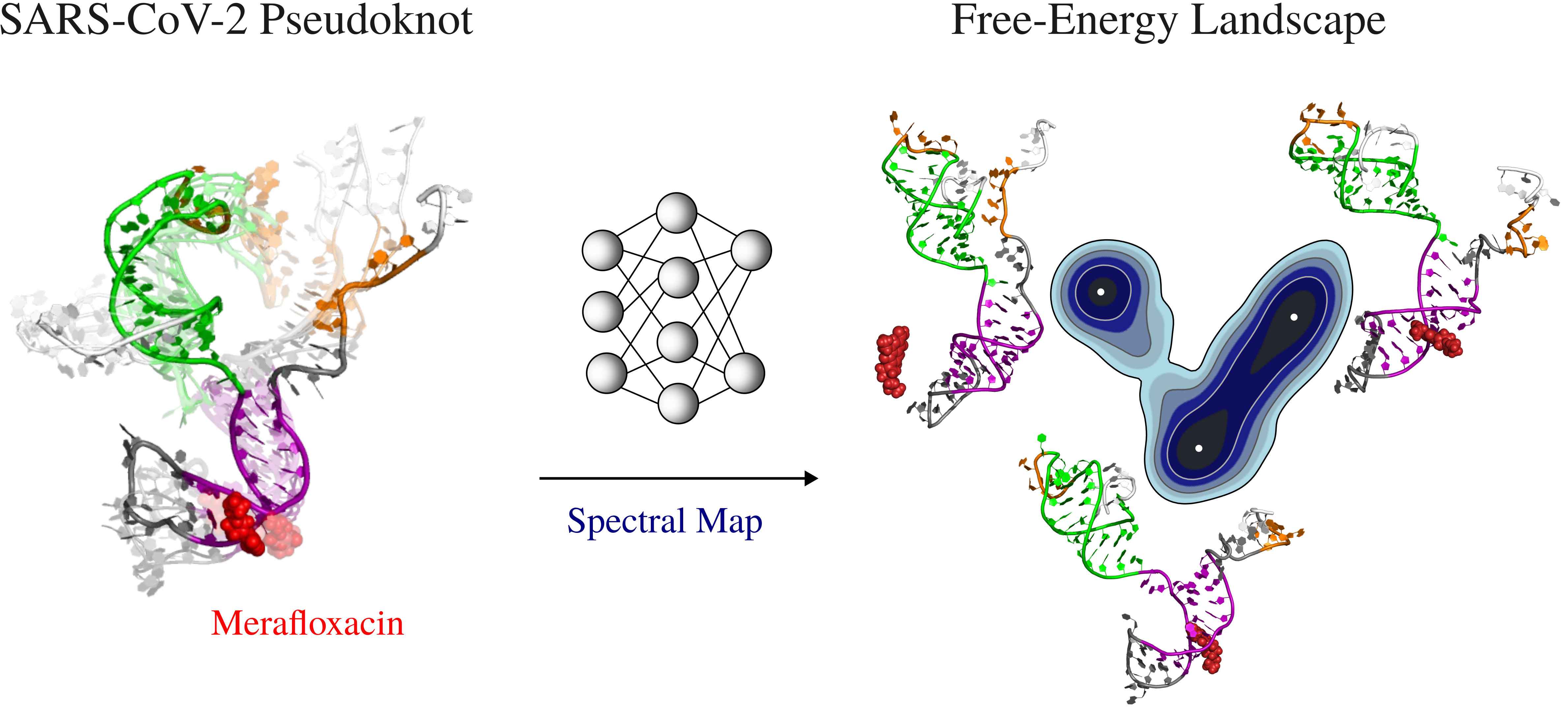}
\end{figure}
In this work, we used molecular dynamics simulations coupled with thermodynamics-inspired machine learning to reveal how ligand binding reshapes the slow conformational dynamics of the SARS-CoV-2 frameshift-stimulatory RNA pseudoknot. Comparing threaded and unthreaded topologies bound to merafloxacin and two analogs in different protonation states, we identify topology-, pose-, and protonation-dependent pathways of ligand-induced conformational changes in the pseudoknot. The resulting free-energy landscapes connect ligand chemistry with rare RNA transitions underlying $-1$ programmed ribosomal frameshifting.

\end{document}